\documentclass[aps,prl,twocolumn,preprintnumbers,10pt,showpacs]{revtex4-1}
\usepackage{amsmath,bm,braket}

\usepackage{graphicx}

\newcommand{\nb}{\bar{n}}

\newcommand{\nbslash}{\rlap{\hspace{0.05em}/}{\bar{n}}}

\newcommand{\Bcal}{\mathcal{B}}

\newcommand{\Pcal}{\overline{\mathcal{P}}}

\allowdisplaybreaks

\begin{document}

\preprint{ZU-TH 18/12}

\title{Transverse parton distribution functions at
  next-to-next-to-leading order: the quark-to-quark case}

\author{Thomas Gehrmann}
\author{Thomas L\"ubbert}
\author{Li Lin Yang}
\affiliation{Institute for Theoretical Physics, University of
  Z\"urich, CH-8057 Z\"urich, Switzerland}


\begin{abstract}
  We present a calculation of the perturbative quark-to-quark
  transverse parton distribution function at next-to-next-to-leading
  order based on a gauge invariant operator definition. We demonstrate
  for the first time that such a definition works beyond the first
  non-trivial order. We extract from our calculation the coefficient
  functions relevant for a next-to-next-to-next-to-leading logarithmic
  $Q_T$ resummation in a large class of processes at hadron colliders.
\end{abstract}

\maketitle

Parton distribution functions (PDFs) describe the momentum
distribution of quarks and gluons inside hadrons, and are essential
inputs for physics program at hadron colliders such as the CERN Large
Hadron Collider (LHC). Their usefulness resides in the factorization
theorems \cite{Collins:1989gx} which separate physics at different
length scales. While PDFs may depend on the 4-momentum of the parton
(the so-called fully-unintegrated PDFs), most physical observables are
only sensitive to the forward component of the parton momentum. To be
more precise, it is convenient to introduce two light-like vectors $n$
and $\nb$ satisfying $n^2=\nb^2=0$ and $n \cdot \nb = 2$, where $n$ is
along the beam direction of the hadron. Any 4-vector $v^\mu$ can then
be decomposed as $v^\mu = \nb \cdot v \, n^\mu/2 + n \cdot v \,
\nb^\mu/2 + v_\perp^\mu$. For an energetic parton along the
$n$-direction, its momentum $q^\mu$ has the hierarchical structure $q
= (\nb \cdot q, n \cdot q, q_\perp) \sim Q \, (1,\lambda^2,\lambda)$,
with $\lambda \ll 1$. For hard interactions, the smaller components $n
\cdot q$ and $q_\perp$ can often be neglected. The resulting PDFs are
referred to as ``collinear PDFs'', which only depend on the fraction
$z = \nb \cdot q / \nb \cdot p$ with $p$ being the hadron momentum.

The collinear PDFs can be defined rigorously in quantum chromodynamics
(QCD) as matrix elements of certain non-local operators
\cite{Collins:1981uw}. For example, the quark PDF is given by
\begin{align}
  \label{eq:phi}
  \phi_{q/N}(z,\mu) &= \int \frac{dt}{2\pi} \, e^{-izt \nb \cdot p}
  \nonumber
  \\
  &\times \braket{N(p) | [\bar{\psi}W](t\nb) \, \frac{\nbslash}{2} \,
    [W^\dagger\psi](0) | N(p)} \, ,
\end{align}
where $\psi$ is the quark field and $W$ is a light-like Wilson line
which renders the operator gauge-invariant. The PDFs are
non-perturbative objects, but their renormalization group (RG)
evolution can be determined perturbatively as long as the
factorization scale $\mu \gg \Lambda_{\text{QCD}}$. Their anomalous
dimension functions (splitting functions) have been computed up to 3
loops \cite{DGLAP3}.

Certain observables, such as the transverse momentum distributions in
the production of the Higgs boson or gauge bosons, however, are
sensitive to the perpendicular component $q_\perp$ of the parton
momentum. One therefore needs to keep $q_\perp$ in the definition of
the relevant PDFs, which are referred to as
transverse-momentum-dependent PDFs (TMDPDFs) or simply transverse
PDFs. Closely related to the transverse PDFs is the
Collins-Soper-Sterman (CSS) formalism of transverse momentum
resummation \cite{Collins:1984kg}, which addresses the divergent
behavior of fixed-order calculations at small transverse momentum. The
CSS formula for the Drell-Yan process can be written in the form
\begin{align}
  \label{eq:css}
  \frac{d\sigma}{dQ^2dQ_T^2dy} &= \frac{4\pi^2\alpha^2}{9Q^2s} \int
  \frac{d^2x_\perp}{(2\pi)^2} \, e^{-iQ_\perp \cdot x_\perp} \sum_q
  e_q^2 \nonumber
  \\
  &\hspace{-5em} \times \exp \left\{ -\int_{\mu_b}^{Q^2}
    \frac{d\bar{\mu}^2}{\bar{\mu}^2} \left[ \ln\frac{Q^2}{\bar{\mu}^2}
      \, A\big(\alpha_s(\bar{\mu})\big) +
      B\big(\alpha_s(\bar{\mu})\big) \right] \right\} \nonumber
  \\
  &\hspace{-5em} \times \Big[ \Pcal_{q/N_1}(z_1,x_T) \,
  \Pcal_{\bar{q}/N_2}(z_2,x_T) + (q \leftrightarrow \bar{q}) \Big] \,
  ,
\end{align}
where $Q^\mu$ is the momentum of the Drell-Yan pair, $x_\perp$ is the
variable conjugate to $Q_\perp$ (also referred to as the ``impact
parameter'' $b$ in the literature), and $\mu_b=b_0/x_T$ with $b_0 =
2e^{-\gamma_E}$. The functions $\Pcal$ can be interpreted as
transverse PDFs. For $x_T^2 \equiv -x_\perp^2 \ll
1/\Lambda_{\text{QCD}}^2$, one can match $\Pcal$ onto the collinear
PDFs
\begin{align*}
  \Pcal_{q/N}(z,x_T) = \sum_i \int_z^1 \frac{d\xi}{\xi} \,
  C_{qi}\big(\xi,\alpha_s(\mu_b)\big) \, \phi_{i/N}(z/\xi,\mu_b)
\end{align*}
with perturbatively calculable coefficient functions $C_{qi}$. These
coefficient functions were determined to the next-to-leading order
(NLO) in \cite{Davies:1984hs, Collins:1984kg}. The general structure
of $C_{ij}$ at NLO was obtained in \cite{deFlorian} including $C_{gi}$
relevant for Higgs production. Previous results beyond NLO were based
on a modification of the CSS formula \cite{Catani:2000vq}, introducing
functions $\mathcal{H}_{ij \leftarrow ab}$, which are related to the
convolutions of $C_{ia}$ and $C_{jb}$ functions. The
next-to-next-to-leading order (NNLO) corrections to $\mathcal{H}$ were
computed \cite{Grazzini} for the Drell-Yan process and for Higgs
production. The method used in these results is to assume the formula
(\ref{eq:css}), and compare its fixed-order expansion with explicit
calculations of the $Q_T$ spectrum. It is however desirable to have an
operator definition of the transverse PDFs, and compute the matching
coefficient functions from the definition.

One may extend the definition (\ref{eq:phi}) to the case of transverse
PDFs as \cite{Becher:2010tm}
\begin{align}
  \label{eq:Bcal}
  \Bcal_{q/N}(z,x_T^2,\mu) &= \int \frac{dt}{2\pi} \, e^{-izt \nb
    \cdot p} \nonumber
  \\
  &\hspace{-5em} \times \braket{N(p) | [\bar{\psi}W](t\nb+x_\perp) \,
    \frac{\nbslash}{2} \, [W^\dagger\psi](0) | N(p)} \, .
\end{align}
This definition, however, is problematic since new divergences
associated with the light-cone propagators arise, which are not
regulated in dimensional regularization. Therefore, one needs to
supplement Eq.~(\ref{eq:Bcal}) with some extra regulator to make it
well-defined. Generically, the result will then depend on this extra
regulator, while in physical observables such dependence will cancel.

Different kinds of regulators have been proposed in the literature.
The original paper \cite{Collins:1981uw} employed a non-light-like
axial gauge. That was employed in subsequent NLO calculations
\cite{NLOcalc}. More recently, Collins introduced a gauge-invariant
definition utilizing non-light-like Wilson lines
\cite{Collins:2011zzd}. An equivalent \cite{Collins:2012uy} definition
has been put forward in \cite{Echevarria:2012qe}. Variants of
Smirnov's analytic regulator \cite{Smirnov:2002pj} were used in
\cite{Becher:2010tm, Becher:2011dz, Chiu:2012ir}. All these approaches
were argued to be valid to all orders in perturbation theory based on
factorization properties. However, explicit calculations were only
carried out at NLO. In this Letter, we report the first NNLO result
for the transverse PDF in the quark-to-quark case, using the regulator
proposed in \cite{Becher:2011dz}. We also extract the NNLO coefficient
function $C_{qq}^{(2)}$, which is the first direct calculation of this
function.

We consider processes where a $q\bar{q}$ pair annihilates into some
color neutral final state $F$: $q(p_1) + \bar{q}(p_2) \to F(Q)$, with
$p_1$ along the $n$-direction and $p_2$ along the $\nb$-direction. We
follow closely the formalism in \cite{Becher:2010tm}, where the
transverse PDF for quarks along the $n$-direction is defined as in
Eq.~(\ref{eq:Bcal}), while the PDF for quarks along the opposite
direction $\bar{\Bcal}_{\bar{q}/N}$ is defined with $n \leftrightarrow
\bar{n}$. To compute the matching functions, we replace the hadron
field $N$ with a quark field and evaluate the matrix element in
Eq.~(\ref{eq:Bcal}). The $\Bcal$ functions require an extra regulator
beyond dimensional regularization, for which we adopt the one
introduced in \cite{Becher:2011dz}, namely, we multiply a factor
$(\nu/n \cdot k)^\alpha$ for each emitted parton with momentum $k$.
The $\Bcal$ functions will contain poles in the analytic regulator
$\alpha$, which however will cancel in the product $\Bcal_{q/q} \,
\bar{\Bcal}_{\bar{q}/\bar{q}}$. A remnant of this regulator dependence is the
collinear anomaly 
\cite{Becher:2010tm},
resulting in a dependence on the hard momentum transfer $Q^2$ in the product,
which can be refactorized as:
\begin{align}
  \label{eq:refac}
  &\left[ \Bcal_{q/q}(z_1,x_T^2,\mu) \,
    \bar{\Bcal}_{\bar{q}/\bar{q}}(z_2,x_T^2,\mu) \right]_{Q^2}
  \\
  &= \left( \frac{x_T^2Q^2}{4e^{-2\gamma_E}}
  \right)^{-F_{q\bar{q}}(L_\perp,\mu)} B_{q/q}(z_1,L_\perp,\mu) \,
  B_{\bar{q}/\bar{q}}(z_2,L_\perp,\mu) \, , \nonumber
\end{align}
where $L_\perp = \ln (x_T^2\mu^2/b_0^2)$. Note that
$B_{q/q}=B_{\bar{q}/\bar{q}}$, we therefore do not distinguish them
anymore. The $B_{i/j}$ functions can be regarded as the
process-independent transverse PDFs, which can be matched onto the
collinear ones for $x_T \ll 1/\Lambda_{\text{QCD}}$ in the form
\begin{align}
  \label{eq:match}
  B_{i/j}(z,L_\perp,\mu) = \sum_k \int_z^1 \frac{d\xi}{\xi} \,
  I_{i/k}(\xi,L_\perp,\mu) \, \phi_{k/j}(z/\xi,\mu) \, .
\end{align}
We define the perturbative expansion of the $I_{i/j}$ functions as
$I_{i/j} = \sum_n (\alpha_s/(4\pi))^n I_{i/j}^{(n)}$, and similarly
for other functions. The second order coefficient $I_{q/q}^{(2)}$ is
the main result in this Letter, while the other combinations of $i$
and $j$ will be presented in a forthcoming article.

At leading order (LO), it is clear that
$\Bcal_{q/q}^{(0)}(z,x_T^2,\mu) =
\bar{\Bcal}_{\bar{q}/\bar{q}}^{(0)}(z,x_T^2,\mu) = \delta(1-z)$. At
NLO, these functions were calculated in \cite{Becher:2010tm} using a
slightly different way of regularization. For our purpose, we need to
recompute them with the regularization scheme of \cite{Becher:2011dz}.
The results read
\begin{align}
  &\Bcal^{(1)}_{q/q}(z,x_T^2,\mu) = C_F \, e^{(\epsilon+\alpha)L_\perp
    - (\epsilon+2\alpha)\gamma_E} \,
  \frac{\Gamma(-\epsilon-\alpha)}{\Gamma(1+\alpha)} \nonumber
  \\
  &\times \left( \frac{\nb \cdot p_1}{\mu} \right)^{\alpha} \left(
    \frac{\nu}{\mu} \right)^{\alpha} (1-z)^{-1+\alpha} \, [ 4z +
  2(1-\epsilon)(1-z)^2 ] , \nonumber
  \\
  &\bar{\Bcal}^{(1)}_{\bar{q}/\bar{q}}(z,x_T^2,\mu) = C_F \,
  e^{\epsilon L_\perp - \epsilon \gamma_E} \, \Gamma(-\epsilon)
  \\
  &\times \left( \frac{\mu}{n \cdot p_2} \right)^{\alpha} \left(
    \frac{\nu}{\mu} \right)^{\alpha} (1-z)^{-1-\alpha} \, [ 4z +
  2(1-\epsilon)(1-z)^2 ] . \nonumber
\end{align}
It is easy to check that the poles in $\alpha$ vanish in the
combination $\delta(1-z_1) \, \bar{\Bcal}^{(1)}_{\bar{q}/\bar{q}}(z_2)
+ \Bcal^{(1)}_{q/q}(z_1) \, \delta(1-z_2)$. The bare
$F_{q\bar{q}}^{(1)}$ and $I_{q/q}^{(1)}$ functions can then be
extracted from this combination following Eqs.~(\ref{eq:refac}) and
(\ref{eq:match}). The remaining poles in the dimensional regulator
$\epsilon=(4-d)/2$ can be renormalized in the $\overline{\text{MS}}$
scheme:
\begin{align}
  \label{eq:ren}
  F_{i\bar{i}}^{\text{bare}}(L_\perp) &= Z^F_{i\bar{i}}(\mu) +
  F_{i\bar{i}}(L_\perp,\mu) \, ,
  \\
  I_{i/j}^{\text{bare}}(z,L_\perp) &= \sum_k \int_z^1 \frac{d\xi}{\xi}
  \, I_{i/k}(\xi,L_\perp,\mu) \, Z^I_{k/j}(z/\xi,L_\perp,\mu) \, .
  \nonumber
\end{align}
Note that for the renormalization at NNLO, we will need
$I_{q/g}^{(1)}$ and $Z^{I,(1)}_{g/q}$, which we also computed.

\begin{figure*}[t]
  \centering
  \begin{tabular}{ccc}
    \includegraphics[width=0.2\textwidth]{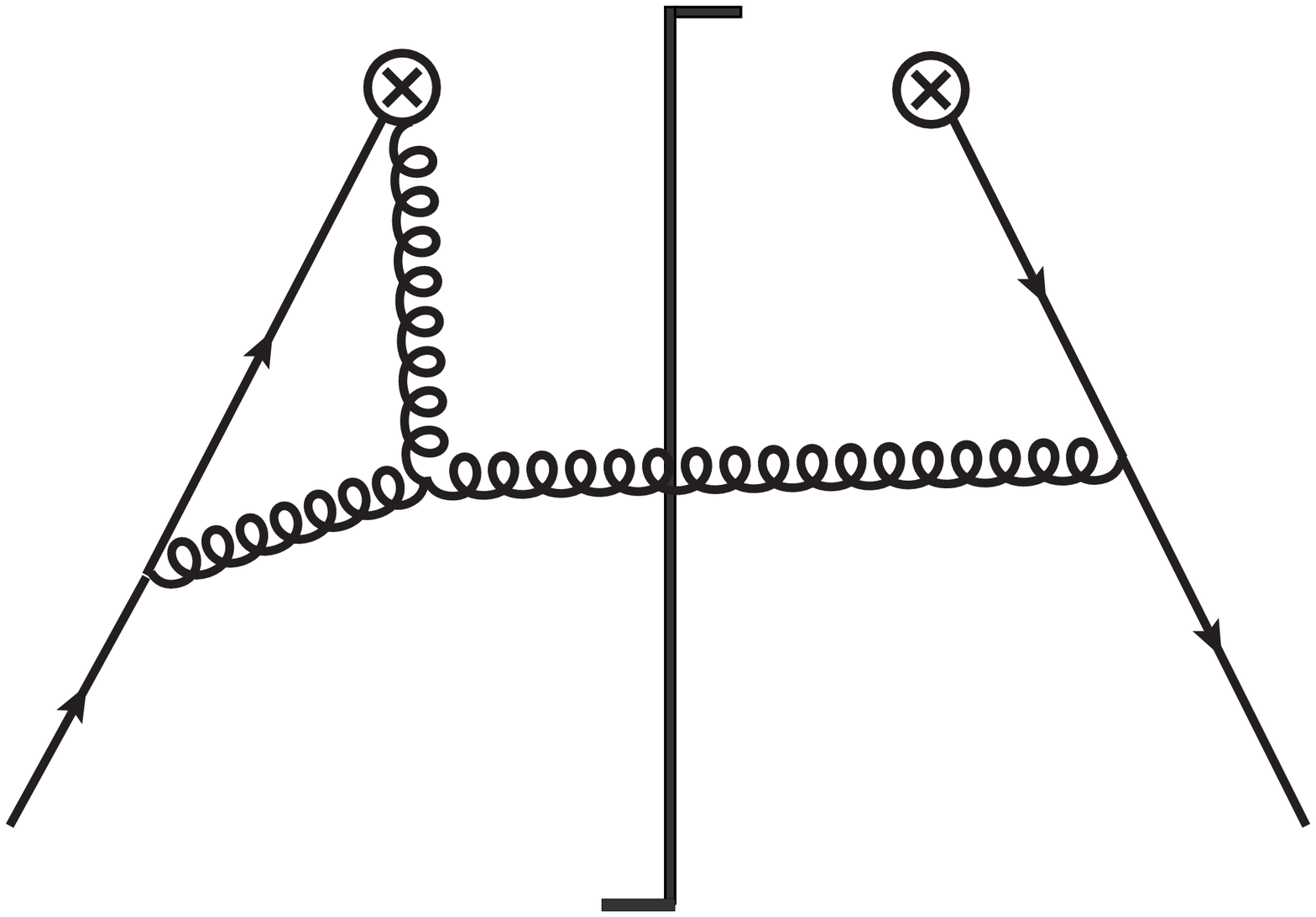}
    &
    \includegraphics[width=0.2\textwidth]{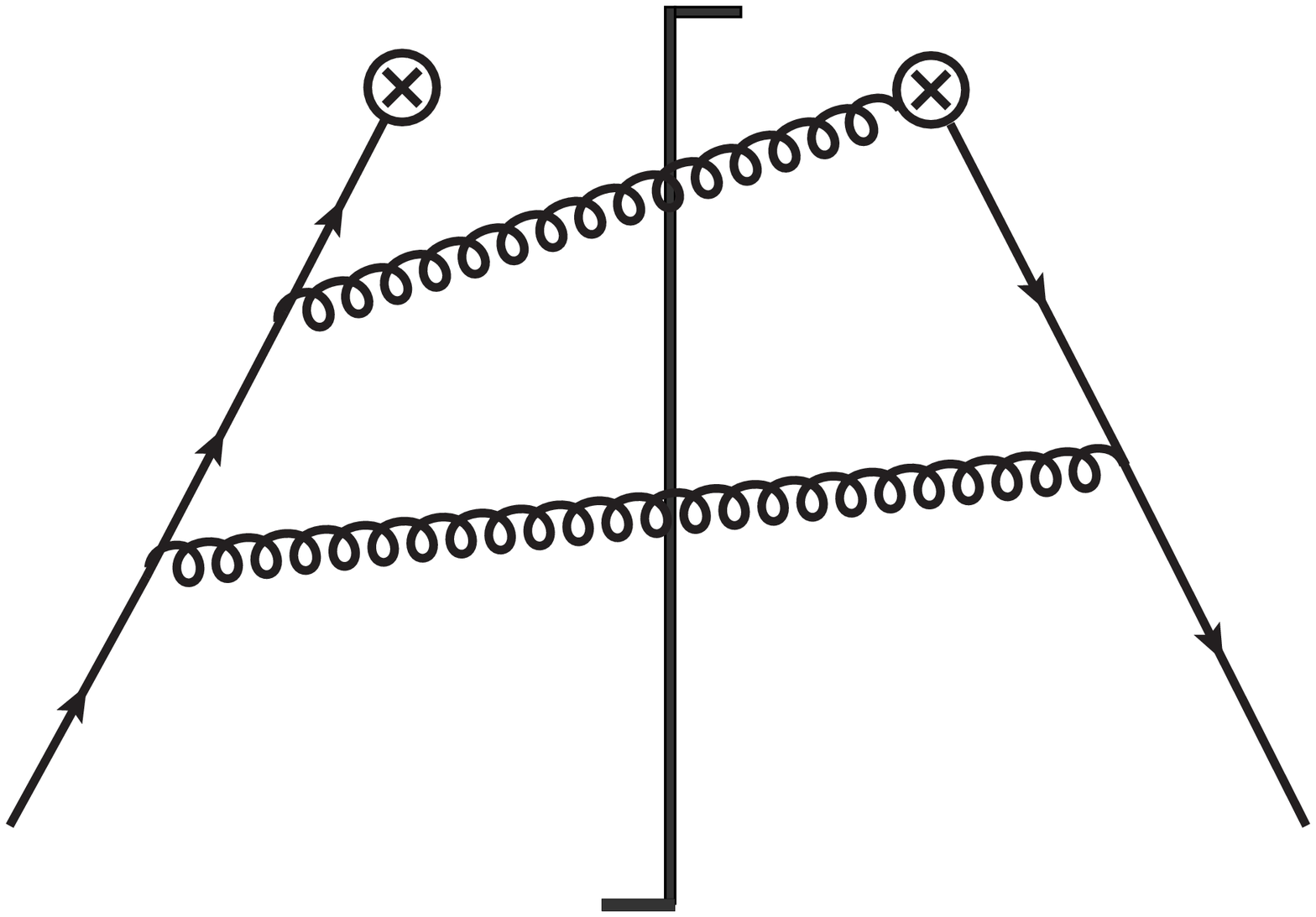}
    &
    \includegraphics[width=0.2\textwidth]{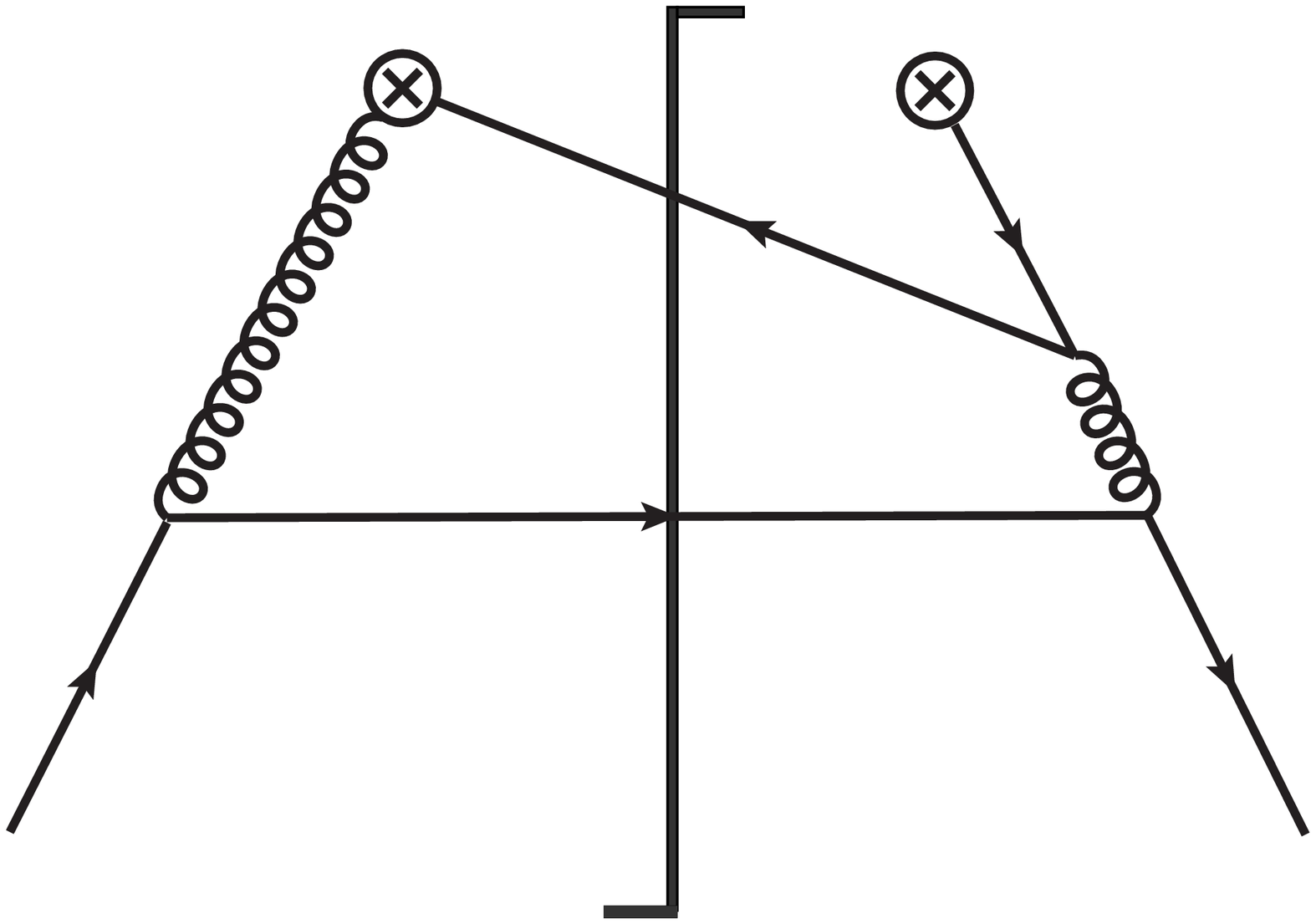}
    \\
    (a) & (b) & (c)
  \end{tabular}
  \caption{Sample Feynman diagrams for our calculation: (a)
    virtual+real; (b) double gluon emission; (c) quark-antiquark pair
    emission.}
  \label{fig:dia}
\end{figure*}

At NNLO, the transverse PDFs receive 3 classes of contributions. In
Figure \ref{fig:dia}, we show a typical Feynman diagram for each of
them: (a) virtual+real diagrams; (b) double gluon emission diagrams;
(c) quark-antiquark pair emission diagrams. The virtual+real diagrams
are relatively easy to calculate. After carrying out the loop
integrals, we encounter familiar integrals which already appeared at
NLO. The virtual+real diagrams contain divergences requiring coupling
constant renormalization, for which we include diagrams with the loop
replaced by a counter-term. The main complication comes from the
double real emission diagrams (b) and (c), where two propagators are
raised to non-integer powers. We managed to reduce them to two-fold
integrals involving hypergeometric functions. We then perform the
remaining integration by a systematic expansion in $\alpha$ and
$\epsilon$. To this end we made extensive use of properties of
hypergeometric functions and the \texttt{HypExp} package
\cite{Huber:2005yg}. The anti-collinear PDF
$\bar{\Bcal}_{\bar{q}/\bar{q}}$ can be obtained similarly, with $n
\leftrightarrow \bar{n}$. Note however that the analytic regulator is
still given by $(\nu/n \cdot k)^\alpha$.

The $\Bcal_{q/q}^{(2)}$ and $\bar{\Bcal}_{\bar{q}/\bar{q}}^{(2)}$
functions contain poles in the analytic regulator $\alpha$. It is
important to verify that these poles cancel in the combination
$\delta(1-z_1) \, \bar{\Bcal}_{\bar{q}/\bar{q}}^{(2)}(z_2) +
\Bcal_{q/q}^{(2)}(z_1) \, \delta(1-z_2) + \Bcal_{q/q}^{(1)}(z_1) \,
\bar{\Bcal}_{\bar{q}/\bar{q}}^{(1)}(z_2)$. We therefore show below the
singular structures in these two functions:
\begin{align}
  &\Bcal_{q/q}^{(2)}(z,L_\perp) = 4C_F \,
  e^{2(\epsilon+\alpha)L_\perp} \left( \frac{\bar{n} \cdot p_1}{\mu}
  \right)^{2\alpha} \left( \frac{\nu}{\mu} \right)^{2\alpha} \nonumber
  \\
  &\times \Bigg\{ \frac{P_{qq}^{(0)}(z)}{\alpha} \left(
    \frac{1}{\epsilon^2} + \zeta_2 \right) -
  \frac{2C_F(1-z)}{\epsilon\,\alpha} + \frac{\delta(1-z)}{\alpha}
  \nonumber
  \\
  &\quad \times \Bigg[ C_F \bigg[ \left( \frac{2}{\alpha} - 3 \right)
  \left( \frac{1}{\epsilon^2} + \zeta_2 \right) - \frac{4}{\epsilon^3}
  + \frac{4\zeta_3}{3} \bigg] + C_A \bigg( \frac{1}{2\epsilon^3}
  \nonumber
  \\
  &\qquad - \frac{67}{36\epsilon} - \frac{101}{27} +
  \frac{23\zeta_3}{6} \bigg) + T_fn_f \bigg( \frac{5}{9\epsilon} +
  \frac{28}{27} \bigg)
  \\
  &\qquad - \frac{\beta_0}{4} \left( \frac{1}{\epsilon^2} + \zeta_2
  \right) \Bigg] \Bigg\} + \Bcal^{(2),\text{ct}}_{q/q}(z,L_\perp) +
  \mathcal{O}(\alpha^0) \, , \nonumber
  \\
  &\bar{\Bcal}_{\bar{q}/\bar{q}}^{(2)}(z,L_\perp) = 4C_F \,
  e^{2\epsilon L_\perp} \left( \frac{\mu}{n \cdot p_2}
  \right)^{2\alpha} \left( \frac{\nu}{\mu} \right)^{2\alpha} \nonumber
  \\
  &\times \Bigg\{ -\frac{P_{qq}^{(0)}(z)}{\alpha} \left(
    \frac{1}{\epsilon^2} + \zeta_2 \right) +
  \frac{2C_F(1-z)}{\epsilon\,\alpha} + \frac{\delta(1-z)}{\alpha}
  \nonumber
  \\
  &\quad \times \Bigg[ C_F \left( \frac{2}{\alpha} + 3 \right) \left(
    \frac{1}{\epsilon^2} + \zeta_2 \right) - C_A \bigg(
  \frac{1}{2\epsilon^3} - \frac{67}{36\epsilon} - \frac{101}{27}
  \nonumber
  \\
  &\qquad + \frac{23\zeta_3}{6} \bigg) - T_fn_f \bigg(
  \frac{5}{9\epsilon} + \frac{28}{27} \bigg) + \frac{\beta_0}{4}
  \left( \frac{1}{\epsilon^2} + \zeta_2 \right) \Bigg] \Bigg\}
  \nonumber
  \\
  & - \Bcal^{(2),\text{ct}}_{q/q}(z,L_\perp) + \mathcal{O}(\alpha^0)
  \, ,
\end{align}
where $P_{qq}^{(0)}(z) = 2C_F[(1+z^2)/(1-z)]_+$, $\beta_0 = 11C_A/3 -
4T_Fn_f/3$. The function $\Bcal^{(2),\text{ct}}_{q/q}$ comes from
$\alpha_s$ renormalization and is given by
\begin{align}
  \Bcal^{(2),\text{ct}}_{q/q}(z,L_\perp) = 4C_F \, e^{\epsilon
    L_\perp} \, \frac{\delta(1-z)}{\alpha} \, \beta_0 \left(
    \frac{1}{\epsilon^2} + \frac{\zeta_2}{2} \right) .
\end{align}
From the above formulae, it is clear that the pole terms with the
color structures $C_FC_A$ and $C_FT_Fn_f$ vanish in the sum $
\delta(1-z_1) \, \bar{\Bcal}_{\bar{q}/\bar{q}}^{(2)}(z_2) +
\Bcal_{q/q}^{(2)}(z_1) \, \delta(1-z_2)$. The remaining singularities
are canceled by the product $\Bcal_{q/q}^{(1)}(z_1) \,
\bar{\Bcal}_{\bar{q}/\bar{q}}^{(1)}(z_2)$.

We are now ready to extract the functions $F_{q\bar{q}}^{(2)}$ and
$I_{q/q}^{(2)}$, following the procedure of Eqs.~(\ref{eq:refac}) and
(\ref{eq:match}) and carrying out the renormalization as in
Eq.~(\ref{eq:ren}). We have checked that the $F_{q\bar{q}}^{(2)}$
function extracted from our calculation agrees with the expression
given in \cite{Becher:2010tm} and that the $I_{q/q}^{(2)}$ function
satisfies the RG equation
\begin{align}
  &\frac{dI_{q/q}(z,L_\perp,\mu)}{d\ln\mu} = \Bigl[
  \Gamma^F_{\text{cusp}}(\alpha_s) L_\perp - 2\gamma^q(\alpha_s)
  \Bigr] I_{q/q}(z,L_\perp,\mu) \nonumber
  \\
  &\hspace{3em} - \sum_k 2 \int_z^1 \frac{d\xi}{\xi} \,
  I_{q/k}(\xi,L_\perp,\mu) \, P_{kq}(z/\xi,\mu) \, ,
\end{align}
where the anomalous dimensions $\Gamma^F_{\text{cusp}}$ and $\gamma^q$
up to 2 loops can be found in \cite{Becher:2010tm}, and the splitting
functions $P_{ij}$ up to 2 loop order can be found in \cite{DGLAP2}.
The finiteness of $I_{q/q}^{(2)}$ and its RG properties demonstrate,
for the first time, that the operator definition for the transverse
PDFs supplemented with the analytic regulator is valid beyond the
first non-trivial order.

We finally give the scale-independent part of the $I_{q/q}^{(2)}$
function, which is the main result of this Letter. It can be written
as
\begin{align}
  &I_{q/q}^{(2)}(z,0) = \delta(1-z) \Bigg[ C_F^2 \frac{5\zeta_4}{4} +
  C_FC_A \bigg( \frac{3032}{81} - \frac{67\zeta_2}{6} \nonumber
  \\
  &- \frac{266\zeta_3}{9} + 5\zeta_4 \bigg) + C_FT_Fn_f \bigg(
  -\frac{832}{81} + \frac{10\zeta_2}{3} + \frac{28\zeta_3}{9} \bigg)
  \Bigg] \nonumber
  \\
  &+ P_{qq}^{(0)}(z) \Bigg[ C_F \bigg( 12\zeta_3 + 4H_0 + \frac{3}{2}
  H_{0,0} + 4H_{0,1,0} + 2H_{0,1,1} \nonumber
  \\
  &- 2H_{1,0,0} + 4H_{1,0,1} + 4H_{1,1,0} \bigg) + C_A \bigg( \zeta_3
  -\frac{202}{27} - \frac{38}{9} H_0 \nonumber
  \\
  & - \frac{11}{6} H_{0,0} - H_{0,0,0} - 2H_{0,1,0} - 2H_{1,0,1} -
  2H_{1,1,0} \bigg) \nonumber
  \\
  &+ T_Fn_f \bigg( \frac{56}{27} + \frac{10}{9} H_0 + \frac{2}{3}
  H_{0,0} \bigg) \Bigg] \nonumber
  \\
  &+ P_{qg}^{(0)}(z) C_F \bigg[ - \frac{68}{27} + \frac{4\zeta_2}{3} +
  \frac{32}{9} H_0 - \frac{4}{3} H_{0,0} + \frac{4}{3} H_{1,0} \bigg]
  \nonumber
  \\
  &+ P_{gq}^{(0)}(z) T_F \bigg[ \frac{86}{27} - \frac{4\zeta_2}{3} -
  \frac{4}{3} H_{1,0} \bigg] \nonumber
  \\
  &+ C_F^2 \bigg[ (2-24z)H_0 + (3+7z)H_{0,0} + 2(1+z)H_{0,0,0}
  \nonumber
  \\
  &+ 2zH_1 + (1-z) \Big( 6\zeta_2 - 22 + 4H_{0,1} + 12H_{1,0} \Big)
  \bigg] \nonumber
  \\
  &+ C_FC_A \Bigg[ (2+10z)H_0 - 4zH_{0,0} - 2zH_1 \nonumber
  \\
  &+ (1-z) \bigg( \frac{44}{3} - 6\zeta_2 - 4H_{1,0} \bigg) \Bigg]
  \nonumber
  \\
  &+ C_FT_F \Bigg[ \frac{-50+38z}{9} + \frac{20+8z}{9} H_0 +
  \frac{2-22z}{3} H_{0,0} \nonumber
  \\
  &+ 4(1+z)H_{0,0,0} \Bigg] - \frac{4}{3} C_FT_Fn_f (1-z) \, ,
\end{align}
where $P_{qg}^{(0)}(z)=2T_F[z^2+(1-z)^2]$,
$P_{gq}^{(0)}(z)=2C_F[1+(1-z)^2]/z$, and $H_{\{m\}} \equiv H(\{m\},z)$
are harmonic polylogarithms introduced in \cite{Remiddi:1999ew}.

The coefficient function $C_{qq}$ relevant for the Drell-Yan process is
related to $I_{q/q}$ by \cite{Becher:2010tm}
\begin{align}
  C_{qq}(z,\alpha_s(\mu_b)) = \left| C_V(-\mu_b^2,\mu_b) \right|
  I_{q/q}(z,0,\alpha_s(\mu_b)) \, ,
\end{align}
where the function $C_V$ is determined from the virtual corrections to
the Drell-Yan process and can be extracted from
\cite{Matsuura:1987wt}. The NNLO expression for $C_{qq}$ reads
\begin{align}
  &C_{qq}^{(2)}(z,\alpha_s(\mu_b)) = I_{q/q}^{(2)}(z,0) + C_F^2 \,
  (1-z) \, (14\zeta_2 - 16) \nonumber
  \\
  &+ C_F \, \delta(1-z) \times \Bigg[ C_F \bigg( \frac{255}{8} -
  19\zeta_2 - 30\zeta_3 + \frac{87\zeta_4}{4} \bigg) \nonumber
  \\
  &\quad + C_A \bigg( -\frac{51157}{648} + \frac{1061\zeta_2}{18} +
  \frac{313\zeta_3}{9} - 8\zeta_4 \bigg) \nonumber
  \\
  &\quad + T_Fn_f \bigg( \frac{4085}{162} - \frac{182\zeta_2}{9} +
  \frac{4\zeta_3}{9} \bigg) \Bigg] \, .
\end{align}
Starting from this expression, we have checked that we can reproduce
the $\mathcal{H}_{q\bar{q} \leftarrow q\bar{q}}^{(2)}$ function in
\cite{Grazzini}. Note that $I_{q/q}$ is universal, while $C_{qq}$ and
$\mathcal{H}_{q\bar{q} \leftarrow q\bar{q}}$ contain both
process-independent and process-dependent parts \cite{Catani:2000vq}.
It is straightforward to compute the $C_{qq}$ and
$\mathcal{H}_{q\bar{q} \leftarrow q\bar{q}}$ functions from our
results up to NNLO for any $q\bar{q}$ initiated process given the
knowledge of the two-loop virtual corrections.

In conclusion, we have calculated the perturbative quark-to-quark
transverse PDF at NNLO based on a gauge invariant operator definition
with an analytic regulator. We demonstrate for the first time that
such a definition works beyond the first non-trivial order. We extract
from our calculation the coefficient functions relevant for a N$^3$LL
$Q_T$ resummation. Our results can be applied to all quark-antiquark
annihilation processes yielding a colorless final state, provided the
NNLO virtual corrections are known. Combined with the recent work
\cite{Zhu:2012ts}, our results could also be applied to the $q\bar{q}
\to t\bar{t}$ process. Our method of calculation can be easily
extended to all parton combinations, which will be presented in a
forthcoming article.

We thank Thomas Becher and Guido Bell for useful discussions. This
work was supported in part by the Schweizer Nationalfonds under grant
200020-141360/1, and by the Research Executive Agency (REA) of the
European Union under the Grant Agreement number PITN-GA-2010-264564
(LHCPhenoNet).

\end{document}